\documentclass{svjour2}                      
\smartqed  
\usepackage{graphicx}
\usepackage{mathptmx}      
\newcommand{\muas}{$\mu$as}
\newcommand{\mas}{m.a.s.}
\newcommand{\degree}{^\circ}
\newcommand{\etal}{et al.}
 \journalname{Experimental Astronomy}
\begin{document}

\title{Gamma Ray Fresnel lenses -- why not?}


\author{G. K. Skinner}


\institute{G. K. Skinner \at
{Centre d'Etude Spatiale des Rayonnements,} \\
{9, avenue du Colonel Roche, 31028,}\\
{Toulouse, France }
              Tel.: +0033-5-61558561\\
              Fax: +0033-5-61556651\\
              \email{skinner@cesr.fr}
}

\date{Received: date / Accepted: date}

\maketitle

\begin{abstract}

Fresnel lenses offer the possibility of concentrating the flux of
X-rays or gamma-rays flux falling on a geometric area of many
square metres onto a focal point which need only be a millimetre
or so in diameter (and which may even be very much smaller). They
can do so with an efficiency that can approach 100\%, and yet they
are easily fabricated and have no special alignment requirements.
Fresnel lenses can offer diffraction-limited angular resolution,
even in a domain where that limit corresponds to less than a micro
second of arc.

Given all these highly desirable attributes, it is natural to ask
why Fresnel gamma ray lenses are not already being used, or at
least why there is not yet any mission that plans to use the
technology. Possible reasons (apart from the obvious one that
nobody thought of doing so) include the narrow bandwidth of simple
Fresnel lenses, their very long focal length, and the problems of
target finding. It is argued that none of these is a `show
stopper' and that this technique should be seriously considered
for nuclear astrophysics. \keywords{Gamma-ray Astronomy \and
Optics}
\end{abstract}

\section{Introduction: Focusing as a phase control problem}

`Gamma-wave' is a convenient term to employ when considering optics
for high energy radiation (gamma-rays, or hard X-rays) in which the
wave-like properties of the radiation need to be taken into account.
We commence with a discussion of the general principles of focussing
`gamma-waves'.

\begin{figure}[h]\centerline{\scalebox{1.0}
{\includegraphics[width=130mm,angle=0.]{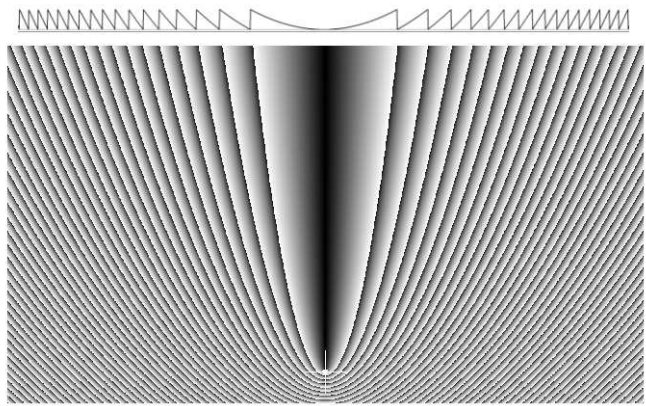}}}
\vspace{-17 mm}
 \caption{
 {The phase at which radiation must be scattered to focus incoming radiation with
  plane horizontal wavefronts.
  At each position the intensity indicates the (relative) phases such that radiation will be focussed
  at the point indicated, with from $0$--$2\pi$ coded  white to
 black. A cross-section is shown above the main figure. In three
  dimensions the iso-phase surfaces are paraboloids of revolution
  about the vertical axis.
 }}
 \label{phase_change}
  \end{figure}

\label{sec:intro} In its simplest form Fermat's principle states
that the path of radiation through an optical system is that which
takes the least time. To be more precise, the time has {\it a
stationary value} with respect to a small deviation in the path.
In this form, Fermat's principle does not take into account the
wave aspect of the radiation -- for example it does not work for
diffraction gratings. However if restated in the form that the
{\it phase} of the radiation at the destination has a stationary
value, then it is applicable quite generally. Thus the various
paths through an optical system must result in the radiation
arriving at the focal point with the same phase, modulo $2\pi$.
This is of course equally obvious if one considers Huygen's
principle and the fact that one wants all wavelets to interfere
constructively.

Suppose that we wish to bring parallel incoming radiation to a
focus by scattering  from elements within an optical system
(`scattering' should  here be thought of as generation of a
Huygen's wavelet; one could equally write `reflection' or
`diffraction'). Figure \ref{phase_change} shows the phase change
necessary as a function of the position of the scattering element
in the case of plane incoming wavefronts (source at infinity).

\begin{figure}[p]
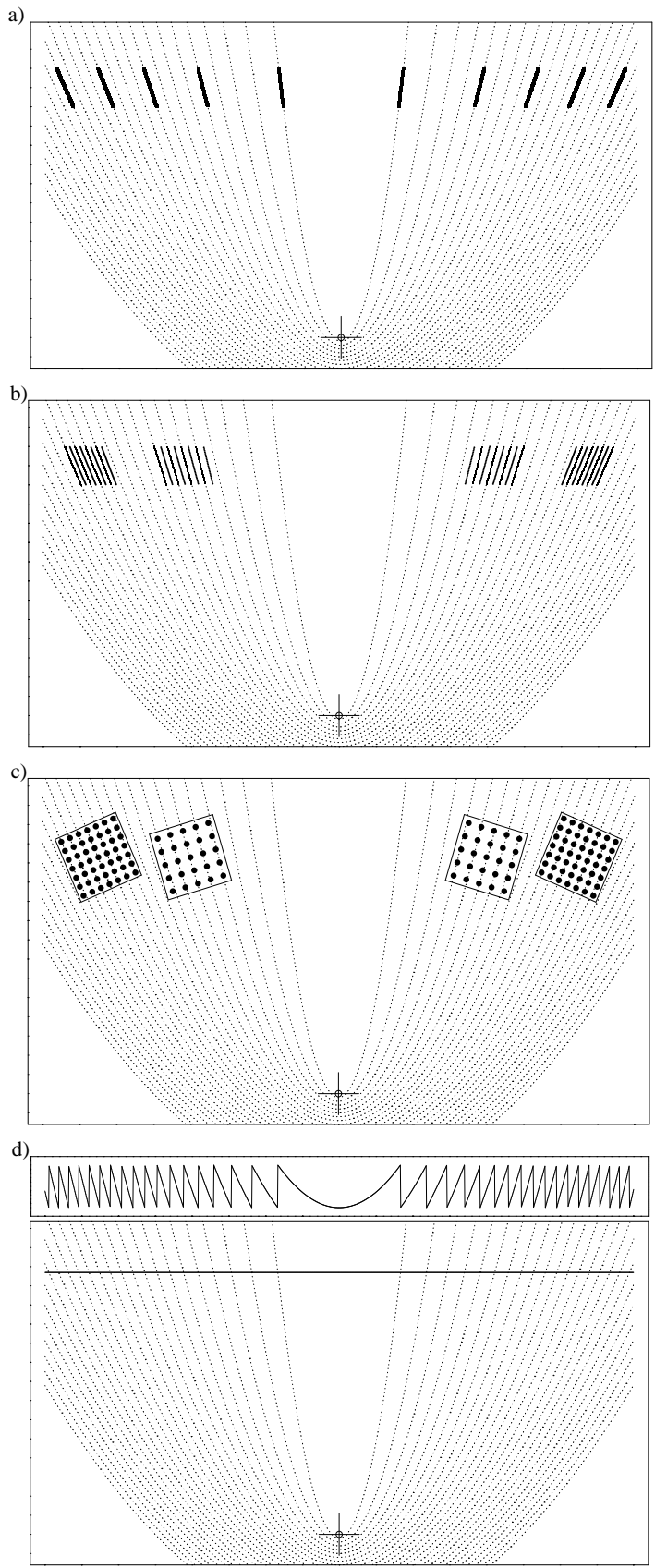

\centerline{a)
\scalebox{1.0}{\includegraphics[width=50mm,angle=-90.]{skinner_bonifacio_fig2a.ps}}}
\vspace{2mm}
\centerline{b)\scalebox{1.0}{\includegraphics[width=50mm,angle=-90.]{skinner_bonifacio_fig2b.ps}}}
\vspace{2mm}
\centerline{c)\scalebox{1.0}{\includegraphics[width=50mm,angle=-90.]{skinner_bonifacio_fig2c.ps}}}
\vspace{2mm}
\centerline{d)\scalebox{1.0}{\includegraphics[width=59mm,angle=-90.]{skinner_bonifacio_fig2d.ps}}}

 \caption{
 The relationship of different focussing systems to isophase contours of
 Figure \ref{phase_change}. (a)~Nested grazing incidence mirrors should
 ideally follow a subset of the isophase surfaces of Figure 1.
 (b)  The layers of multilayer mirrors follow also follow such surfaces.
 (c) In the case of a Laue lens, crystals planes approximate iosphase surfaces.
 (d) A cut along the line shown gives the phase change needed in a  planar focussing device.
 }
 \label{system_examples}
  \end{figure}

Often each scattering takes place with the same phase change
(diffraction by identical atoms, reflection from surfaces ...). In
this case the scatterers need to be distributed on iso-phase
contours, which form a system of nested paraboloids. Figure
\ref{system_examples} illustrates how grazing incidence mirrors,
multilayer optics, and Laue diffraction `lenses' use respectively
reflecting surfaces, alternating layers, and planes of atoms to
aim to populate parts of iso-phase surfaces.

In each case only an approximation to the ideal is achieved. For
Laue lenses, the planes of atoms are flat and equispaced and so
cannot perfectly follow the surfaces. In this case, and for
grazing incidence or multilayer optic, practical tolerances are
such that the scattering is coherent only very small regions. Lack
of coherence means that on larger scales intensities are added,
not amplitudes.

\subsection{Changing the phase of a gamma-wave}

\label{phase_section}  The requirements for focussing can be
 met exactly if the phase of the gamma-wave can be controlled.
Figure \ref{system_examples}(d) indicates how the phase change
needs to vary across the surface of a planar focussing component
in order to concentrate plane incoming wave onto a point. How does
one change the phase of a gamma-wave? It turns out to be
surprisingly easy \cite{Paper1}.

It is well known that grazing incidence reflection relies on the
fact that for X-rays and gamma-rays refractive indices are
slightly less than unity. Conventionally one writes %
$    n= 1-\delta$,  where $\delta$ is small and positive. Thus the
phase of X/$\gamma$-ray radiation will be changed by any material
that it passes through. It is useful to define
$t_{2\pi}=\lambda/\delta$, which is the thickness of material
which leads to a phase change of $2\pi$ with respect to the phase
{\it in vacuo}. This is of course the largest phase change ever
needed. Table \ref{eg_table} gives an indication of orders of
magnitude of $t_{2\pi}$ for some example materials and photon
energies, as well as of the transmission losses associated with
this amount of material. Over a wide range of materials and
energies, dimensions are convenient for manufacture and losses are
very low.

\begin{table}[h]
 {\caption{Examples of the thickness of material necessary to
change the phase of X/$\gamma$-ray radiation by $2\pi$, and the
corresponding absorption loss.  Low atomic number materials show
the least absorbtion. Gold is an example of a high density
material which though having poor transmission at low energies
minimises the lens thickness. \label{eg_table}}}
\begin{center}
\begin{tabular}{lcccccccc}
\hline \vspace{-2.5mm} \\
  & \multicolumn{2}{c}{5 keV}   & \hspace{3mm} & \multicolumn{2}{c}{100 keV}& \hspace{3mm} & \multicolumn{2}{c}{500 keV}    \\
Material      &  $t_{2\pi}$  & Absorption   & & $t_{2\pi}$  & Absorption   &  &$t_{2\pi}$  & Absorption      \\
              &    $\mu$m    & \%           & &  $\mu$m     & \%           &  &  $ mm  $   & \%              \\
\hline
              &              &              &  &            &              &  &           &                  \\
Polycarbonate &  23          &     6.3      &  &   470      &      0.9     &  &     2.4      &   2.6         \\
Silicon       &  13          &     51       &  &   260     &      0.9     &  &     1.3      &   2.7            \\
Gold          &   2.1        &     93       &  &    39      &      32      &  &     0.19     &   5.2           \\
  \hline
\end{tabular}
\end{center}
\end{table}

\subsection{An example  Phase Fresnel Lens for gamma-rays}

A consequence of the possibility of modulating the phase of a
gamma-wave with a convenient thickness of material is that one can
make a lens for focussing gamma-rays simply by giving a disk of
material a thickness profile with the form shown in Figure
\ref{system_examples}(d). This is essentially a Fresnel lens, but
as the term is often applied to a simple lens whose thickness is
stepped without consideration to maintaining phase, we term it a
Phase Fresnel Lens, or PFL.

As an example, such a lens for 500 keV gamma-rays might consist of
a 5 m diameter disk of (say) aluminium with a thickness varying
between $t_{min}=0.25$ mm (to provide a substrate) and
$t_{max}=t_{min}+t_{2\pi}=1.4$ mm. The absorption losses would be
1.6\%. We shall consider below the example of such a lens in which
the finest pitch of the pattern (at the periphery) is $\sim$1 mm.

\section{Advantages of Phase Fresnel Lenses}

Phase Fresnel Lenses seem to offer major advantages for gamma-ray
astronomy.

\paragraph{Angular resolution: }
The diffraction-limited angular resolution of a PFL of diameter
$d$ at wavelength $\lambda$ is given by the usual formula for a
circular aperture, $\theta_D=1.22 \lambda/d$.  In appropriate
circumstances there is no reason why resolution close to this
limit should not be achieved. For a 5 m diameter lens at 500 keV
the diffraction limit is 0.12 {\bf micro} seconds of arc (\muas).
Sub micro arc second resolution is exactly what is needed to image
the space-time around a black hole - a specific objective of
NASA's `Beyond Einstein' program.

\paragraph{Simple manufacture: }
Because the refractive index is very close to unity, the
manufacturing tolerances necessary in the construction of a PFL
are often relatively lax. For an aluminium lens working at 500
keV, $\lambda/40$ optical precision requires only 30 micron
accuracy. In such circumstances Gamma-ray lens `polishing' can be
done with regular machine-shop tools!

 \label{tolerance_section}

\paragraph{Collecting area: }
Because of the low transmission losses and the large geometrical
collecting area possible with PFLs, the effective area of a
PFL-based telescope can be enormous. Even allowing for absorption,
for reasonable detector efficiency (50\%), and for {\it rms}
wavefront errors corresponding to $\lambda/10$, the effective area
of the example 5 m diameter lens would be 65000 cm$^2$.

\paragraph{Low aberrations and non-critical alignment: }
With the typical dimensions which are being considered here, a PFL
is an extreme example of a high aperture-ratio ($f$-number), thin
lens. Consequently geometrical aberrations are low (in fact
entirely negligible for any conceivable size focal plane detector)
and any tilt of the lens with respect to the viewing direction has
little effect (tilts of $\sim 1\degree $ can be tolerated). In
addition the depth of field is relatively large.

\begin{table}[ht] {\caption{Examples of the focal length $f$ of
a PFL having diameter, $d= 5$ m for various values of finest pitch
$p$ and photon energy. The aspect ratio is indicative ($t_{2\pi}/p$)
and supposes a moderately dense material ($\rho\sim$10).The first
two lines illustrate energy/pitch/focal-length combinations for
which PFL aspect ratios are probably impracticable and for which
Laue diffraction (Fig. 1c) is more appropriate. \label{f_p_table}}}
\begin{center}
\begin{tabular}{lcccc}\hline
Pitch $p$      & Energy (keV)    &   Aspect ratio & Focal length $f$ & Notes \\
\hline
                &                &                &                &               \\
0.5 nm          &  100           &  $1.3\times 10^5$         &  100 m         &   Typical atomic spacing      \\
0.5 nm          &  1000          &  $1.3\times 10^6$        &  1 km          &      " ~~~~~~~ " ~~~~~~~"      \\
200 nm          &   5            &     16         &   2 km         &   Note 1   \\
25 microns      &  100           &     2.5        &  5000 km       &   Note 2      \\
1.0 mm          &  500           &     0.32       & 10$^6$ km      &    Example used here     \\
  \hline
\end{tabular}
\end{center}
  {\small { Note 1 Same pitch as the Chandra  HEG  grating
  \cite{heg}} \\
   \small {Note 2  Same aspect ratio as the Chandra HEG Diffraction
  grating}  }
\end{table}

\begin{figure}[h]\centerline{\scalebox{1.0} {\includegraphics[width=60mm,angle=-90.]{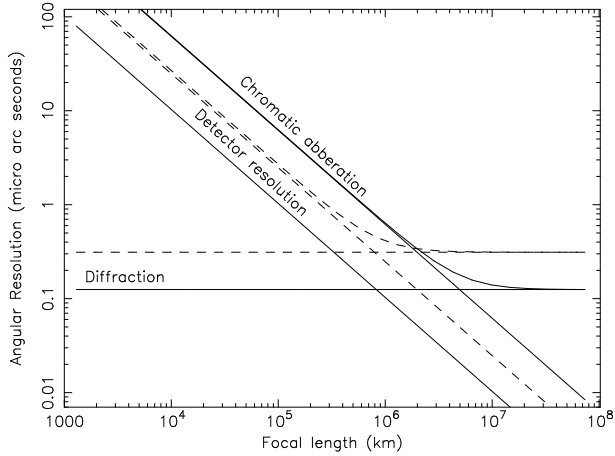}}}
 \caption{
 Continuous lines: the contributions to the angular resolution of a 5 m diameter PFL for 500 keV
 gamma-rays. That for chromatic abberation assumes a 2 keV band.  Dashed lines show the corresponding values for a 2 m diameter PFL (the detector is of course unchanged).
 In each case the curves show the combination of the three
 components.
 }
 \label{angres_plot}
  \end{figure}

\section{The downside}

For a given lens diameter and $\lambda$ the focal length of a
Fresnel lens is given by $f=pd/2\lambda$. Table \ref{f_p_table}
gives some example values for a $5$ m diameter lens. It can be
seen that unless one reduces $p$ to the atomic scale (as in Laue
lenses), to the nanometric scale (as in multilayer mirrors), or at
least to that entailing micromachining, then focal lengths are
{\it very} long, particularly for high energies.

Minimising the focal length of a PFL by adopting very small $p$ is
difficult because the necessary thickness (depth) of the structure
remains that discussed in Section \ref{phase_section} and the
`aspect ratio' (pitch/depth) of the required profile becomes very
high. Also, although it was argued in Section \ref
{tolerance_section} that tolerances on the thickness are easily
achieved, radial tolerances need to be  $\ll p$ if diffraction
limited focussing if not to be compromised.

However, there are  reasons for not trying to minimise $f$. The
diffraction limited resolution will not be obtained if the
detector does not have adequate spatial resolution or if chromatic
aberration is too important. Combining the above expression for
the focal length in terms of $p$ with $\theta_D=1.22 \lambda/d$
shows that the physical dimension of the diffraction spot  is
simply $0.66p$, independent of $\lambda$. A detector with spatial
resolution $\Delta x$ which is of this order of magnitude or worse
will cause blurring on an angular scale $\Delta x/f$. Thus any PFL
system with $\Delta x >~p$, or equivalently $ f< \Delta x\
d/2\lambda$, will be limited by detector resolution, not by
diffraction. In the case of chromatic abberation,  a spectral band
of width $\Delta \lambda$  leads to a blurring $\theta_C = 0.15
(d/f)(\Delta\lambda/\lambda)$. Thus both detector resolution
problems and those of chromatic aberration ease with increasing
$f$.

Figure  \ref{angres_plot} shows the effect on the contributions to
the angular resolution of changing the design focal length of an
example Fresnel lens having $d= 5$ m and working at 500 keV. For the
baseline design it is supposed that the detector resolution is
limited by the track length of the electron receiving the energy of
the incoming to $\Delta x\sim 0.5$ mm. Similarly, the spectral band
is assumed to be no narrower than the typical resolution of a
Germanium detector,  $\Delta E \sim 2$ keV at 500 keV. It can be
seen that unless $f >\sim 10^6 $km, then chromatic abberation will
limit the angular resolution to no better than about one \muas.

Thus the two main problems which need to be addressed are
chromatic aberration and long focal length.

\subsection{Chromatic aberration}

The problems caused by chromatic aberration include both the loss
of angular resolution due to blurring of the image and the
decrease in sensitivity to which this blurring leads through the
necessity of collecting the signal from a larger, and hence higher
background, region of the detector. We have seen that the former
effect is mimimised by using long focal lengths. The latter bears
some discussion.

Suppose one accepts that chromatic abberation is inevitable.
Obviously for mono-energetic radiation there is not a problem. For
broadband emission, one will properly focus only the radiation in a
small fraction of the spectrum. With an energy resolving detector,
radiation outside a band $\Delta E$ can be ignored. One can then
pose the question - how wide should $\Delta E$ be for the best
sensitivity? Suppose the sensitivity is limited by noise on the
detector background in a focal spot of diameter $d_d$. We may take
the detector background to be proportional to $\Delta E\ d_d^2$. If
$d_d$ is dictated by chromatic aberration then it will be
proportional to $\Delta E$. For a continuum source, so will the
signal $S$. Consequently, assuming that the dominant source of noise
is statistical background fluctuations, the signal-to-noise ratio
$S/B^{1/2}$ is proportional to $\Delta E^{-{1/2}}$. One thus reaches
the counter-intuitive conclusion that, in these circumstances, the
best signal-to-noise ratio is obtained by using a band as narrow as
the detector energy resolution will allow. This is of course also
best for angular resolution.

Using a narrow energy band, the sensitivity to a broadband signal
will of course not be as good as if a wider bandpass had been
possible without compromising the focusing, but given the large
collecting area and the tiny background in a small focal spot {\it
and} in a narrow energy range, it can be amazingly good. The
narrow line sensitivity of our example 5 m, 500 keV lens could be
$<10^{-8}$ photons cm$^{-2}$ s$^{-1}$.

Of course, with an energy resolving detector one can use data from
energies on either side of the design energy $E$ of a PFL, but in
general away from the optimum energy the sensitivity and angular
resolution are not so supremely good. Diffraction limited focussing
does occur also at harmonics of $E$, but the efficiency is low
(ignoring differences in absorption, the response at $2E$ is 4.5\%
of that at $E$).

  In many circumstances it may be desirable to accept a lower
sensitivity in order to gain information at other energies. The
available surface area can then be divided into zones tuned to
different energies, as is often proposed for Laue lenses. The
zoning of the surface can be radial, azimuthal, or into separate
lenses with independent focal spots. For many configurations the
result will be that a broadband spectrum will be sampled at a
limited number of relatively narrow bands -- a situation which
often has to be accepted in radio astronomy.

\begin{figure}[h]\centerline{\scalebox{1.0} {\includegraphics[width=50mm,angle=-90.]
{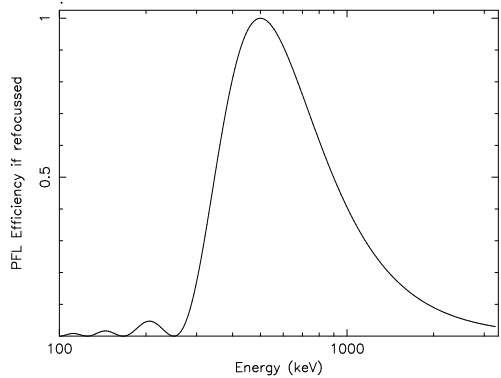}}}
 \caption{
The efficiency of a 500 keV PFL if the detector distance is
adjusted to match the focal length for each energy (note - the
corresponding curve in reference \cite{Paper1} is slightly
incorrect).
 }
 \label{refocus_plot}
  \end{figure}

Another possibility is to take advantage of the fact that a PFL
works over a broad band of energies provided the instrument is
refocused by moving the detector to the appropriate focal plane.
Figure \ref{refocus_plot} shows that the $\Delta E/E$ obtainable
in this way is $\sim$~1. The time available for each measure will
of course be shorter, so in this case too the sensitivity at each
energy will be poorer.

Achromatic modifications of PFLs have been proposed (see
\cite{skinner_achrom} and references therein). These rely on
combinations of diffractive and refractive components. In
practical cases the refractive component has to be stepped to
avoid excessive absorption. Such systems hold considerable
promise, though because of the stepping the best performance is
obtained only for a comb of energies and absorption losses tend to
be significant.

 It is worth noting that Laue lenses are just as subject
to chromatic aberration as PFLs. Over a wide energy band,
radiation falling on part of a PFL will be diffracted, though not
in the ideal direction. This is directly analogous to Laue
diffraction by a mosaic crystal - where again all energies within
a certain band are diffracted but with some dispersion in the
diffracted direction. Assuming small angles, in each case one
finds $\Delta\theta/\theta\sim-\Delta E/E$. The differences are
(i) crystal pass bands much narrower and (ii) Laue lenses have
imaging properties which are much poorer. Consequently the
chromaticity is often insignificant compared with other
limitations. Interestingly, if the crystals could be composed of
crystallites (or laminae) with the same orientation but having
different crystal plane spacing it could be avoided. Such a system
would be analogous to a graded spacing multilayer mirror but in
Laue geometry in place of Bragg.

\subsection{The focal length}

It has been argued above that to get the best performance, long
focal lengths are inevitable. Given that formation flying
technology is necessary for {\it any } focal length above 10--20
m, what are the constraints?

First it is worth noting that the idea of formation flying
satellites with a scale of 10$^6$km is not unprecedented. The joint
ESA-NASA Lisa gravitational wave mission will involve 3 spacecraft
in a triangular formation with $5\times 10^6$ km between each pair.
The major difference between the formation flying needed for LISA
and that for  a for gamma-ray  PFL mission is that the LISA
formation rotates with respect to inertial space in such a way that
each spacecraft follows the local gravity field.

Suppose that we have a one spacecraft with a PFL and another
carrying the corresponding detector. If  the vector positions of
the spacecraft are  $\mathbf{v_1}$ and $\mathbf{v_2}$, the
direction of the separation vector
$\mathbf{f}=\mathbf{v_1}-\mathbf{v_2}$ is the pointing direction
of the instrument. For it to be constant we must have
$\mathbf{\ddot{v}_1}=\mathbf{\ddot{v}_2}$. If the first spacecraft
is in a free orbit then its acceleration $\mathbf{\ddot{v}_1}$ is
provided by the gravitational field at $\mathbf{v_1}$. In general
the gravitational field at $\mathbf{v_2}$ will be different and so
cannot correspond to the acceleration $\mathbf{\ddot{v}_2}$. To
overcome the gravity gradient a station-keeping force must be
supplied to one of the spacecraft by a constantly firing thruster.

Gravity gradients close to the earth are too large for a long
focal length configuration to be feasible . However a study by the
NASA IMDC \cite{imdc} has shown that for a solar orbit, drifting
away from the earth, the thrusts necessary are within the range of
available ion engines and that the fuel requirements, although
large, are not prohibitive. Other aspects of such a mission were
found to present no overriding problems.

The IMDC study set out only to establish feasibility and the orbit
selected is not necessarily optimal. It may be that  there that
there are advantages in having one of the spacecraft relatively
close to the earth-moon system, whose gravitational field could,
with judicious planning, be used to minimise fuel requirements.

To repoint a million km long telescope at a new target will
inevitably take an appreciable time. The IMDC study considered
observations of several weeks, with a significant part of the
mission time, and the majority of the fuel, used for target
changing. It was noted that the efficiency and fuel requirements
could be much improved with two detector (or two lens) spacecraft.

\subsection{Other considerations: Field of view and target finding}

The field of view (fov) over which a gamma-ray PFL yields good
imaging can be very large and geometric aberrations are never
important. In practice the fov will be limited by detector size.
Both pointing accuracy and knowledge of target positions must be
good enough to ensure that the target image falls within the the
part of the image plane recorded by the detector. With a focal
length of 10$^6$km, a 1 m detector gives a fov of only 200 \muas.
For comparison, the FGS points HST to an accuracy of 10 milli arc
seconds (\mas), 50 times worse .

Existing astrometrical measurements do not provide target
locations of the necessary precision. The Hipparchos catalog
achieved $\sim$1 \mas accuracy, though uncertainties in proper
motion mean that it is no longer valid at this level. Successor
missions DIVA and FAME will offer improvements for relatively
bright stars. Other projected optical astrometry projects are
expected to offer major progress in timescales shorter than any
likely launch of a gamma-ray PFL mission. ESA's Gaia (scheduled to
be launched in 2010) will measure 300000 stars to 4 \muas.
Although further into the future and less certain, NASA's OBSS and
SIM PlanetQuest missions plans to push the precision down to close
to 1 \muas.

Optical astrometry relies heavily on centroiding and care is
necessary with AGN, in which the nucleus is generally embedded in
surrounding emission. Radio measurements provide a surer means of
astrometry for AGN and offer precisions which are as good as, and
currently better than, those possible in other bands. At present
the International Celestial Reference system is primarily realised
by the International Celestial Reference Frame of 212
extragalactic radio sources with {\it rms} position uncertainties
of 100 to 500 \muas.


As well as knowing the location of the source, the vector
$\mathbf{f}$ above, defining the separation and the orientation of
the axis of the axis defined by the two spacecraft must be
measured and controlled. The problem has been studied in the
context of the MAXIM mission, leading to proposals of solutions
involving 'super star trackers' and/or super-conducting gyros
\cite{gendreau}. Given the very long focal length required for a
PFL, GPS-like surveying using radio delay/phase measurements to
perform can provide  information on the positions of the
spacecraft in 3-d space and hence constraints on the vector
$\mathbf{f}$, valuable in obtaining an attitude. A baseline
limited by the size of the earth would probably not be adequate,
so beacons or transponders on several further spacecraft would be
desirable. The reference spacecraft need not necessarily be
dedicated to the PFL mission -- beacons on multiple, well
distributed, spacecraft such as those of the LISA mission would be
particularly valuable\footnote {It is interesting to note that if
a total of at least $n\geq7$ stations is involved, measurement of
the distances and velocities between all pairs formally provides
enough information to solve for the $6n$ coordinates locating them
all in momentum space, using no reference other than the structure
of the gravitation potential of the solar system. The formation
itself becomes an inertial sensor!}.

\section*{Conclusions}

\label{conclusions}
With their extremely good imaging properties,
with the exceptionally large effective area possible from a simple
device and with the very low background associated with their
compact focus, PFLs would probably be the universal means of
focussing gamma-rays were it not for the disadvantages discussed
above.

It has been argued above that in appropriate circumstances these
disadvantages can all be overcome and that although an ambitious
mission would be needed, there are no `show-stoppers'.
Particularly where the ultimate in angular resolution or in point
source sensitivity in a relatively narrow energy band is required,
PFLs hold enormous potential in the era of gamma-wave astronomy.


\begin{thebibliography}{}
%
%

\bibitem{Paper1}
Skinner, G.K. `` Diffractive/refractive optics for high energy
astronomy -I : Gamma-ray Fresnel Lenses '' A\&A 375, 691 (2001),
``-II : Variations on the theme'' A\&A 383, 352 (2002)

\bibitem{heg}
Canizares, C. R., \etal\ ``The Chandra High-Energy Transmission
Grating'', PASP, 117 836 (2005)

\bibitem{skinner_achrom}
Skinner, G.K. `` Design and imaging performance of achromatic
diffractive-refractive x-ray and gamma-ray lenses '' Appl. Optics,
43, 4845  (2004)

\bibitem{gendreau}
Gendreau, K. C., \etal\ ``Requirements and options for a stable
inertial reference frame for a 100-micro-arcsecond imaging
telescope'' Proc SPIE 4852, 685 (2002)

\bibitem{gendreau}
Phillips, J.D., \etal\ ``Metrology and pointing for astronomical
interferometers'' Proc SPIE 5491, 320 (2004)

\bibitem{imdc}
Krizmanic, J., \etal\ ``Formation flying for a Fresnel lens
observatory mission'' This volume.



\end{thebibliography}


\end{document}